\documentclass[a4paper,11pt,twocolumn,unpublished]{quantumarticle}
\pdfoutput=1

\usepackage{amsmath,amssymb}

\usepackage[T1]{fontenc}  %
\usepackage{graphicx}
\usepackage{bm}
\usepackage{booktabs}
\usepackage{algpseudocode}  %
\usepackage{xcolor}
\usepackage{tikz}
\usetikzlibrary{quantikz2}  %
\usepackage{braket}         %
\usepackage{enumitem}       %

\newcounter{algenv}
\renewcommand{\thealgenv}{\arabic{algenv}}
\newenvironment{algfloat}[1]{%
  \begin{figure}[t]\small\refstepcounter{algenv}%
  \noindent\rule{\columnwidth}{0.9pt}\par\vspace{2pt}%
  \noindent\textbf{Algorithm~\thealgenv:}~#1\par\vspace{1pt}%
  \noindent\rule{\columnwidth}{0.4pt}\par\vspace{2pt}}{%
  \par\vspace{1pt}\noindent\rule{\columnwidth}{0.9pt}\end{figure}}
\usepackage[colorlinks=true,allcolors=blue]{hyperref}

\newcommand{\dnew}{\textcolor{blue}{\footnotesize$\blacktriangleright$}\,}

\newcommand{\Z}{\mathbb{Z}}

\newcommand{\R}{\mathbb{R}}
\newcommand{\C}{\mathbb{C}}

\newcommand{\ztr}{\Z[\sqrt{2+\sqrt2}]}
\newcommand{\Tcount}{T\text{-count}}

\newcommand{\dd}{d_\diamond}
\newcommand{\norm}[1]{\lVert #1 \rVert}

\begin{document}

\title{Approximate synthesis of general single-qubit unitaries over the Clifford+$\sqrt{T}$ gate set}

\author{Mathias Weiden}
\email{mtweiden@berkeley.edu}
\affiliation{University of California, Berkeley, California, USA}
\affiliation{Lawrence Berkeley National Laboratory, Berkeley, California, USA}

\author{Jae Won Kim}
\affiliation{University of California, Berkeley, California, USA}

\author{Justin Kalloor}
\affiliation{University of California, Berkeley, California, USA}
\affiliation{Lawrence Berkeley National Laboratory, Berkeley, California, USA}

\author{John Kubiatowicz}
\affiliation{University of California, Berkeley, California, USA}

\author{Costin Iancu}
\affiliation{Lawrence Berkeley National Laboratory, Berkeley, California, USA}

\date{September 14, 2026}

\begin{abstract}
    For the standard Clifford+$T$ gate set, \emph{deterministic, ancilla-free} synthesis now attains the minimal $\Tcount$ for general single-qubit unitaries~\cite{morisaki2025optimalancillafreeclifford+t}.
    The $\sqrt{T}$ gate rotates by half the angle of $T$, generating a finer lattice of implementable operations.
    Access to this magic state is expected to lower the cost of deterministic and ancilla-free synthesis of general single-qubit unitaries~\cite{kliuchnikov2023shorterquantumcircuitsa}, but no direct Clifford+$\sqrt{T}$ algorithm existed for this case.
    We provide one by extending the integer lattice-point enumeration method of Morisaki~\textit{et~al.}
    We adopt a resource state cost model based on the magic-state catalysis approach of Gidney and Fowler~\cite{gidney2019efficientmagicstate}.
    On Haar-random targets synthesized to precisions ranging from $\varepsilon=10^{-3}$ to $10^{-8}$, the cost of Clifford+$\sqrt{T}$ circuits scales as $2.4\log_2(1/\varepsilon)$ compared to $3.0\log_2(1/\varepsilon)$ for $\Tcount$-optimal Clifford+$T$ circuits.
    Once a one-time catalyst state is amortized, the Clifford+$\sqrt{T}$ circuits are never costlier than their Clifford+$T$ counterparts.
\end{abstract}

\maketitle

\section{Introduction}

The cost of a fault-tolerant quantum circuit is dominated by its non-Clifford gates.
Clifford gates are comparatively inexpensive.
The standard universal fault-tolerant gate set is Clifford+$T$, whose non-Clifford generator $T$ is a $\pi/4$ rotation about the $Z$ axis.
Because the $T$ gate is most commonly realized by injecting a magic state, distilled~\cite{bravyi2005universalquantumcomputation} or cultivated~\cite{gidney2024magicstatecultivation} at a cost far above that of any Clifford operation, the $\Tcount$ is the standard measure of cost for a sequence of quantum gates.

\begin{figure}[t]
\centering
\includegraphics[width=\columnwidth]{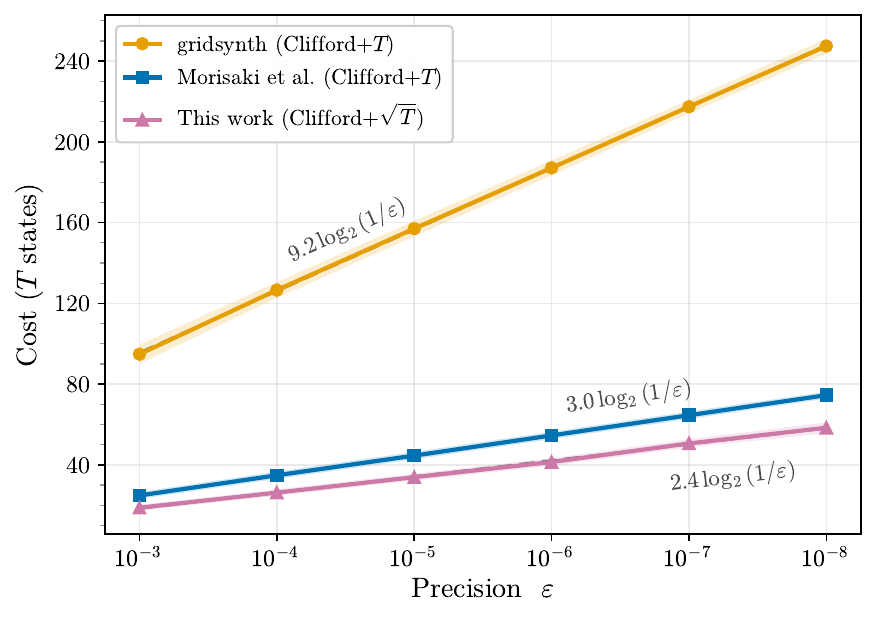}
\caption{
    Non-Clifford cost of synthesizing general single-qubit unitaries versus precision $\varepsilon$. Each point represents the mean of $500$ Haar-random targets per $\varepsilon$.
    Shaded regions represent $\pm1$ standard deviation.
    Dashed lines are fitted $C\log_2(1/\varepsilon)$ scalings (details in Sec.~\ref{sec:results}).
    Prior high-precision methods approximate a general unitary through three independently synthesized single-axis rotations.
    Here \texttt{gridsynth}~\cite{ross2016optimalancillafreeclifford+t} costs ${\sim}9\log_2(1/\varepsilon)$ $T$ states.
    Direct synthesis at optimal $\Tcount$~\cite{morisaki2025optimalancillafreeclifford+t} lowers this to $3.0\log_2(1/\varepsilon)$.
    The direct Clifford+$\sqrt{T}$ synthesis introduced in this work lowers it further, to $2.4\log_2(1/\varepsilon)$ $T$ states under the resource state cost model of Sec.~\ref{sec:cost}.
}
\label{fig:cost-headline}
\end{figure}

\begin{figure*}[t]
\centering
\includegraphics[width=\textwidth]{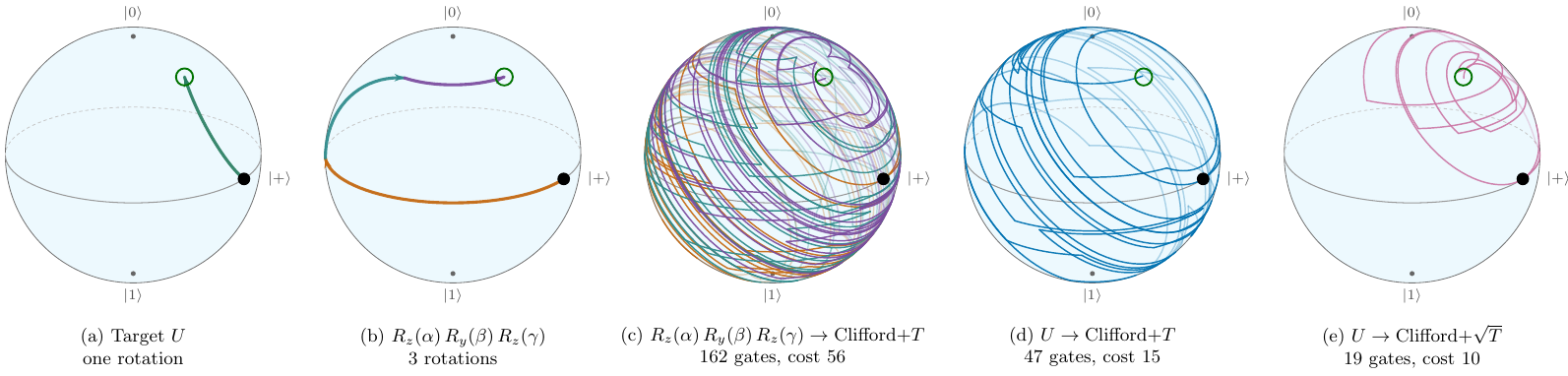}
\caption{
    Five methods of approximating the same unitary.
    Here we illustrate a target unitary $U$ applied to a state $\ket{+}$ and trace its synthesized paths on the Bloch sphere.
    A precision of $\varepsilon = 10^{-2}$ is used in each case.
    The length of the traced trajectory reflects the cost of the resulting circuits.
    (a)~The continuous operation to be approximated: a single rotation carrying the state (black dot) to the target (green ring).
    (b)~The same operation decomposed into three single-axis rotations.
    (c)~The indirect route taken by prior high-precision methods synthesizes each rotation independently.
    The colors of the tracks correspond to the similarly colored segments in (b).
    (d)~Synthesizing $U$ directly into Clifford+$T$~\cite{morisaki2025optimalancillafreeclifford+t} shortens the walk substantially.
    (e)~Direct Clifford+$\sqrt{T}$ synthesis, the contribution of this work, traces less trajectory still.
}
\label{fig:bloch-traj}
\end{figure*}

For the Clifford+$T$ gate set, optimal single-qubit ancilla-free synthesis is solved.
Ross and Selinger showed how to synthesize $z$-rotations with optimal $\Tcount$ up to a factoring oracle~\cite{ross2016optimalancillafreeclifford+t}.
The algorithm due to Morisaki~\textit{et al.}~\cite{morisaki2025optimalancillafreeclifford+t} attains the optimal $\Tcount$ for arbitrary single-qubit unitaries with no factoring assumption.
Unitaries implementable at a given $\Tcount$ are embedded in a discrete lattice.
The algorithm enumerates lattice points closely aligned with a non-lattice target vector.
This process is repeated on successively finer meshes, until a suitable lattice point (i.e.\ one that is within some distance threshold $\varepsilon$ away) is found.

The $\sqrt{T}$ gate rotates about $Z$ by $\pi/8$, half the angle of $T$.
Adjoining it to the Clifford group produces a denser mesh of reachable unitaries, so a target can often be approximated with fewer non-Clifford rotations than its optimal Clifford+$T$ circuit requires.
We do not assume that $\sqrt{T}$ states can be efficiently prepared directly.
They are instead realized by the magic-state catalysis of Gidney and Fowler~\cite{gidney2019efficientmagicstate}, which produces two $\sqrt{T}$ states by consuming five $T$ states.
Half of the time a $T$ state is needed to correct the $\sqrt{T}$ injection, meaning a $\sqrt{T}$ gate costs three $T$ states in expectation.
This method requires a reusable $\sqrt{T}$ catalyst state with a small one-time preparation cost (Sec.~\ref{sec:cost}).
This work establishes that at this cost, synthesis using the finer lattice afforded by the $\sqrt{T}$ gate results in cheaper gate sequences.

\emph{Exact} synthesis, expressing a unitary that is already representable over Clifford+$\sqrt{T}$ as a $\sqrt{T}$-count-minimal circuit, is solved~\cite{forest2015exactsynthesissinglequbit}.
\emph{Approximate} synthesis, where the entries of an arbitrary target $V$ are rounded to the ring within diamond distance $\dd(U,V)<\varepsilon$ at low cost, was addressed for $z$-rotations by Kliuchnikov, Bocharov, Roetteler and Yard~\cite{kliuchnikov2015frameworkapproximatingqubit} and for general unitaries by Kliuchnikov~\textit{et al.}~\cite{kliuchnikov2023shorterquantumcircuitsa}.
Their algorithm enumerates integer points of the gate-set lattice~\cite{lenstra1983integerprogrammingfixed}, but they apply it to the target indirectly.
A general unitary is first split by Euler decomposition into two $z$-rotation approximations and one magnitude approximation.
Each sub-problem is then solved by lattice search.
This baseline reduction is deterministic and uses no ancilla.
Its lowest-cost variants lower the \emph{expected} cost further. Approximations can be \emph{mixed} by classically sampling each solution from an ensemble of lower-precision circuits so that the averaged channel meets a higher precision than its members~\cite{campbell2017shortergatesequences}.
Approximations can also be extended by an ancilla-and-measurement fallback~\cite{paetznick2014repeatuntilsuccessnondeterministicdecomposition, bocharov2015efficientsynthesisprobabilistic}.
Mixing consumes no extra quantum resources, but its $\varepsilon$-accuracy is a property of the sampled ensemble, realized across repeated runs rather than in any single execution.

In every variant the general unitary is approximated indirectly, through its constituent rotations.
A \emph{direct} construction, one that synthesizes the target unitary as a whole rather than through those rotations, deterministically and without ancillas, has so far been demonstrated for Clifford+$T$~\cite{morisaki2025optimalancillafreeclifford+t} but not Clifford+$\sqrt{T}$.
Fig.~\ref{fig:bloch-traj} illustrates these routes on the Bloch sphere.

The contribution of this work is a direct, deterministic, ancilla-free approximate-synthesis algorithm for the Clifford+$\sqrt{T}$ gate set targeting general single-qubit unitaries (Secs.~\ref{sec:problem}--\ref{sec:algorithm}).
Our algorithm produces a single fixed sequence of single-qubit gates that approximates the target to within a diamond distance of $\varepsilon$.
The algorithm extends the lattice method of Morisaki~\textit{et al.}~\cite{morisaki2025optimalancillafreeclifford+t} from $\Z[\omega]$ to $\Z[\zeta]$, where $\omega=e^{2\pi i / 8}$ is the eighth root of unity and $\zeta=e^{2\pi i/16}$ is the sixteenth root of unity.
It remains numerically faithful at practical runtimes down to $\varepsilon=10^{-8}$, and is released as the open-source library \texttt{cyclosynth}\footnote{\url{https://github.com/mtweiden/cyclosynth}}.

On Haar-random targets, under the magic-state cost model motivated in Sec.~\ref{sec:cost}, the non-Clifford cost of our Clifford+$\sqrt{T}$ circuits grows as $2.4\log_2(1/\varepsilon)$, against $3.0\log_2(1/\varepsilon)$ for the $\Tcount$-optimal Clifford+$T$ circuits of the same targets (Fig.~\ref{fig:cost-headline}), a $20\%$ reduction.
Per target, the $\sqrt{T}$ circuits are cheaper in over $99\%$ of cases and, once a one-time catalyst is amortized, never costlier (Sec.~\ref{sec:results}).
The optimality argument of~\cite{morisaki2025optimalancillafreeclifford+t} rests on a relation between $\Tcount$ and lde that is specific to $\Z[\omega]$.
We do not establish such a relation for $\Z[\zeta]$.

\subsection{Main results}
\label{sec:main-results}

Our main results are as follows:

\emph{A direct, deterministic, ancilla-free approximate synthesis algorithm for general single-qubit unitaries over Clifford+$\sqrt{T}$} (Algorithm~\ref{alg:sqrtt}), guaranteeing $\dd(U,V)<\varepsilon$ with gate sequences that are never costlier than the optimal Clifford+$T$ solution.
On Haar-random targets, the non-Clifford cost of our circuits grows as $2.4\log_2(1/\varepsilon)$, versus $3.0\log_2(1/\varepsilon)$ for the $\Tcount$-optimal Clifford+$T$ circuits of the same targets (Fig.~\ref{fig:cost-headline}).

\emph{\texttt{cyclosynth}}, an open-source implementation of the algorithm that synthesizes at precisions down to $\varepsilon=10^{-8}$ in seconds (Table~\ref{tab:runtime}).
On $z$-rotations, general Clifford+$T$ synthesis never needs more $T$ gates than \texttt{gridsynth} (Sec.~\ref{sec:results}).

\begin{figure*}[ht]
\centering
\begin{quantikz}[row sep={0.60cm,between origins}, column sep=0.34cm]
  \lstick{$\ket{+}$} &  & \ctrl{2} &  &  &  &  &  &  &  & \ctrl{3} &  & \ctrl{3} &  &  & \ctrl{2} &  & \rstick{$\sqrt{T}\ket{+}$} \\
  \lstick{$\ket{+}$} &  & \targ{} & \ctrl{2} &  & \targ{} & \gate{T^\dagger} & \targ{} &  &  &  &  &  &  & \ctrl{1} & \targ{} &  & \rstick{$\sqrt{T}\ket{+}$} \\
  \lstick{$\sqrt{T}\ket{+}$} & \gate{X} & \targ{} &  & \ctrl{1} & \targ{} & \gate{T^\dagger} & \targ{} &  &  &  &  &  &  & \control{} & \targ{} & \gate{X} & \rstick{$\sqrt{T}\ket{+}$} \\
  \setwiretype{n} & \lstick{$T\ket{+}$} & \setwiretype{q} & \targ{} & \targ{} & \ctrl{-2} & \gate{T} & \ctrl{-2} & \gate{H} & \gate{S} & \targ{} & \gate{T} & \targ{} & \gate{H} & \meter{}\vcw{-1} & \setwiretype{n} &  &
\end{quantikz}
\caption{
    Catalyzed $\sqrt{T}$-state preparation circuit, following the construction of Gidney and Fowler~\cite{gidney2019efficientmagicstate}.
    A recyclable catalyst $\sqrt{T}$ state must first be initialized before other $\sqrt{T}$ states can be generated.
    Executing this circuit consumes $5$ $T$ states to produce $2$ new $\sqrt{T}$ states.
    Each $\sqrt{T}$ state consumed via state injection requires a $T$-gate correction $50\%$ of the time.
    In expectation, executing $2$ $\sqrt{T}$ gates therefore costs $6$ $T$ states: a single $\sqrt{T}$ gate costs $3$ $T$ states.
    Odd powers of $\sqrt{T}^d$ also have a cost of $3$ $T$ states.
    For example: $\sqrt{T}^3 = \sqrt{T}T = \sqrt{T}^\dagger S$  and injection results in the adjoint case half of the time.
}
\label{fig:catalyst-circuit}
\end{figure*}
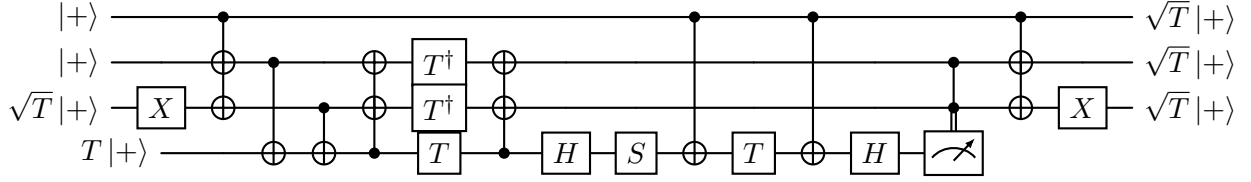

We assume targets are single-qubit unitaries, that error is measured in diamond distance, and circuits are single fixed gate sequences. 
Ancillas and measurements are used only in preparing the magic states that implement $T$ and $\sqrt{T}$ gates (Fig.~\ref{fig:catalyst-circuit}).
We also assume that Clifford gates have negligible cost.
The cost of a gate sequence is the number of $T$ states required to implement it.
A single $T$ gate costs one $T$ state.
Each $\sqrt{T}$ gate implemented via the magic-state catalysis method of Gidney and Fowler~\cite{gidney2019efficientmagicstate} costs three $T$ states in expectation (after the one-time catalyst preparation is amortized, see Sec.~\ref{sec:cost}).
Our conclusions are robust against this exact $\sqrt{T}$ cost.
The advantages of synthesizing to the Clifford+$\sqrt{T}$ gate set persist for any price below the crossover ${\sim}4$ $T$ states (Sec.~\ref{sec:discussion}).
Our empirical claims cover $\varepsilon\in[10^{-3},10^{-8}]$.
We prove no cost-optimality for the Clifford+$\sqrt{T}$ circuits.

\section{Background}
\label{sec:background}

\subsection{Exact and approximate gate synthesis}

Single-qubit synthesis takes an arbitrary $2\times2$ unitary $V$ and returns a sequence of gates from a fixed finite set whose product approximates $V$.
We measure the approximation error by the diamond distance $\dd(U,V)$~\cite{kliuchnikov2023shorterquantumcircuitsa}, which for single-qubit unitaries has a closed form (Appendix~\ref{app:diamond}); approximate synthesis to precision $\varepsilon$ requires $\dd(U,V)<\varepsilon$.

The gate set consists of the Clifford generators $\langle H,S\rangle$ together with one of the non-Clifford generators:
\begin{equation*}
\begin{gathered}
  H=\tfrac{1}{\sqrt2}\begin{pmatrix}1&1\\1&-1\end{pmatrix},\qquad
  S=\begin{pmatrix}1&0\\0&i\end{pmatrix},\\[4pt]
  T=\begin{pmatrix}1&0\\0&e^{i\pi/4}\end{pmatrix},\qquad
  \sqrt{T}=\begin{pmatrix}1&0\\0&e^{i\pi/8}\end{pmatrix}.
\end{gathered}
\end{equation*}
Since $T=\sqrt{T}^2$, every Clifford+$T$ circuit is also a Clifford+$\sqrt{T}$ circuit, and the larger set reaches more unitaries per unit depth.

Synthesis proceeds in two stages.
In \emph{exact} synthesis, a given target can be expressed exactly in the specified gate set.
The goal is to return a lowest cost gate sequence which implements it.
This problem is solved for both considered gate sets~\cite{kliuchnikov2012fastefficientexact,forest2015exactsynthesissinglequbit}.
\emph{Approximate} synthesis instead considers an arbitrary target unitary.
The target must first be rounded to a nearby exactly-implementable unitary and then synthesized using exact methods.
The approximate case is unsolved for general Clifford+$\sqrt{T}$ and is the subject of this paper.

Throughout, synthesis is assumed to be \emph{ancilla-free}.
The output of such a synthesis algorithm is a gate sequence which consists only of single-qubit operations.
Ancillas are used only in preparing the magic states that implement $T$ and $\sqrt{T}$ gates.

\subsection{Implementable unitaries and rings}
\label{sec:implementable}

Up to global phase, a single-qubit unitary is determined by its first column $(u_1,u_2)$,
\begin{equation}
  U = \begin{pmatrix} u_1 & -\bar{u}_2 \\ u_2 & \;\;\bar{u}_1 \end{pmatrix},
  \qquad |u_1|^2 + |u_2|^2 = 1 .
  \label{eq:param}
\end{equation}
A unitary is exactly implementable precisely when $u_1$ and $u_2$ are cyclotomic integers scaled by a power of $\sqrt2$,
\begin{equation}
  u_1, u_2 \,\in\, \tfrac{1}{\sqrt2^{\,k}}\,\mathcal{R},
  \label{eq:lde}
\end{equation}
with $\mathcal{R}=\Z[\omega]$ ($\omega=e^{2\pi i/8}$) for Clifford+$T$~\cite{kliuchnikov2012fastefficientexact} and $\mathcal{R}=\Z[\zeta]$ ($\zeta=e^{2\pi i/16}$) for Clifford+$\sqrt{T}$~\cite{forest2015exactsynthesissinglequbit}.
A cyclotomic integer is an integer combination of powers of the root of unity.
For $\Z[\omega]$ there are four coefficients, for $\Z[\zeta]$ there are eight.
Halving the rotation angle thus doubles the number of integer coordinates per ring element, and with it the dimensionality of the search in Sec.~\ref{sec:problem}.

The smallest exponent $k$ for which Eq.~\eqref{eq:lde} holds is the least denominator exponent (lde).
It is a measure of depth, since a larger $k$ admits a finer mesh of implementable unitaries.
An implementable unitary is a tuple $(u_1,u_2)$ at some scale $k$.
Exact synthesis maps such a tuple to a gate sequence, whereas approximate synthesis must find a tuple whose unitary lies within $\varepsilon$ of the target.
The non-Clifford cost of the resulting circuit grows linearly with $k$~\cite{kliuchnikov2012fastefficientexact,forest2015exactsynthesissinglequbit}, so a deeper $k$ yields a finer mesh at a proportionally higher cost.

\subsection{Resource state cost model}
\label{sec:cost}

We measure the cost of a gate sequence by counting the expected number of $T$ resource state injections required:
\begin{equation}
  \mathrm{cost}(U) = n_T + 3\,n_{\sqrt{T}}.
  \label{eq:cost}
\end{equation}
    A $\sqrt{T}$ gate costs three $T$ states and a $T$ gate costs a single $T$ state (Fig.~\ref{fig:catalyst-circuit}). 
    Gate sequences are decomposed into \emph{syllables} of the form $C^\prime \sqrt{T}^m C$ for some Cliffords $C^\prime, C \in \langle H, S \rangle$ and $m \in \{1, 2, 3\}$.
    The count $n_T$ is the number of syllables with $m=2$, and $n_{\sqrt{T}}$ the number with $m \in \{1, 3\}$.

The construction consumes a $\sqrt{T}$ \emph{catalyst} state that is reused across all injections.
Gidney and Fowler prepare the analogous $T$ catalyst of their factory only to the error rate of the states it processes~\cite{gidney2019efficientmagicstate}.
We conservatively prepare the $\sqrt{T}$ catalyst to $\varepsilon^2$, at a one-time startup cost of $3\log_2(1/\varepsilon^2)=6\log_2(1/\varepsilon)$ $T$ states.
At $\varepsilon=10^{-8}$, where access to $\sqrt{T}$ gates saves ${\sim}22\%$ in non-Clifford cost, the construction breaks even against this startup cost after ${\sim}10$ synthesized operations.
For the many rotations typical of fault-tolerant workloads, the startup is negligible.

Finally, our search runs the Clifford+$T$ synthesizer concurrently and floors the $\sqrt{T}$ result by it, so $\mathrm{cost}(U_{\sqrt T})\le\mathrm{cost}(U_T)$ for every target by construction (Algorithm~\ref{alg:sqrtt}).

\section{Approximate synthesis as a lattice problem}
\label{sec:problem}

To solve the approximate synthesis problem, first fix a scale $k$ and determine whether there is a tuple $(u_1,u_2)/\sqrt2^{\,k}$ whose associated unitary (Eq.~\eqref{eq:param}) lies within $\varepsilon$ of the target, $\dd(U,V)<\varepsilon$.
The ring-basis coordinates of $u_1$ and $u_2$ form an integer vector $x$.
For Clifford+$T$, $x=(a_1,b_1,c_1,d_1,a_2,b_2,c_2,d_2)$ encodes $u_1=a_1+b_1\omega+c_1\omega^2+d_1\omega^3$ and $u_2$ likewise (Appendix~\ref{app:bilinear} treats both rings).
In these coordinates the two requirements, that $U$ be unitary and that $U$ be close to $V$, become geometric conditions on the integer vector $x$.

Unitarity, $|u_1|^2+|u_2|^2=2^k$, splits into a norm shell that places $x$ on a sphere and a set of bilinear equations, quadratic in their coordinates:
\begin{align}
  \text{norm shell:}\quad & \norm{x}^2 = 2^k, \label{eq:normshell}\\
  \text{bilinear:}\quad & \Phi_j(x) = 0. \label{eq:bilinear-omega}
\end{align}
The Clifford+$T$ case has one bilinear equation and Clifford+$\sqrt{T}$ has three (Table~\ref{tab:doubling}).
Appendix~\ref{app:bilinear} derives these forms and verifies that an integer $x$ satisfies Eqs.~\eqref{eq:normshell}--\eqref{eq:bilinear-omega} exactly when it encodes an implementable unitary.

Distance to the target, $\dd(U,V)\le\varepsilon$, is a third geometric condition on $x$
\begin{equation}
  (\hat y \cdot x)^2 \ge \tau(k,\varepsilon),
  \qquad \tau(k,\varepsilon) = 2^k\,(1-\varepsilon^2),
  \label{eq:cap}
\end{equation}
where $\hat y$ is the target column $v=(V_{11},V_{21})$ carried into the lattice coordinates by the embedding of Appendix~\ref{app:embedding}.
Appendix~\ref{app:diamond} reduces the diamond distance to this overlap.
This \emph{cap} condition tightens around the target direction as $\varepsilon\to0$.

An implementable unitary at scale $k$ therefore exists exactly when the lattice $\Z^{2r}$ contains a point that lies on the norm shell, on the bilinear surfaces, and inside the cap (Eqs.~\eqref{eq:normshell}--\eqref{eq:cap}).
Fig.~\ref{fig:cap-geometry} sketches this geometry.
Exact synthesis then realizes the corresponding tuple as a gate sequence whose non-Clifford count grows with $k$~\cite{kliuchnikov2012fastefficientexact,forest2015exactsynthesissinglequbit}.
This is a closest-vector problem.
Morisaki~\textit{et al.} solve it by Lenstra-style enumeration of the lattice points in a convex body~\cite{morisaki2025optimalancillafreeclifford+t,lenstra1983integerprogrammingfixed}; we realize that enumeration by LLL reduction followed by Schnorr--Euchner enumeration~\cite{lenstra1982factoringpolynomialsrational,schnorr1994latticebasisreduction} (Sec.~\ref{sec:algorithm}).

\definecolor{mplblue}{HTML}{1F77B4}
\definecolor{mplorange}{HTML}{FF7F0E}
\definecolor{mplgreen}{HTML}{2CA02C}
\definecolor{mplred}{HTML}{D62728}
\definecolor{mplpurple}{RGB}{148,103,189}

\begin{figure}[t]
\centering
\begin{tikzpicture}[scale=2.8]
  \def\ta{52}        %
  \def\dc{0.90}      %
  \pgfmathsetmacro{\hw}{acos(\dc)}  %
  \def\s{0.141421}   %
  \fill[mplblue!13] (\ta-\hw:1) arc[start angle=\ta-\hw, end angle=\ta+\hw, radius=1] -- cycle;
  \foreach \i in {-9,...,9} \foreach \j in {-8,...,8}
    \fill[gray!48] (\i*\s,\j*\s) circle (0.55pt);
  \draw[thick] (0,0) circle (1);
  \foreach \a/\b in {1/7,7/1,-1/7,-7/1,1/-7,7/-1,-1/-7,-7/-1}
    \fill[gray!70] (\a*\s,\b*\s) circle (0.8pt);
  \foreach \a/\b in {-5/5,-5/-5,5/-5}
    \fill[black!78] (\a*\s,\b*\s) circle (1.05pt);
  \draw[mplblue!60,very thick] (\ta-\hw:1) -- (\ta+\hw:1);
  \draw[mplblue!45, thin] (0,0) -- (\ta-\hw:1);
  \draw[mplblue!45, thin] (0,0) -- (\ta+\hw:1);
  \draw[mplblue!85, thin, {Latex[length=4pt]}-{Latex[length=4pt]}]
    (\ta-\hw:0.62) arc[start angle=\ta-\hw, end angle=\ta, radius=0.62];
  \node[mplblue!85, inner sep=1pt] at (\ta-\hw/2:0.74) {\Large $\varepsilon$};
  \draw[mplorange!60,ultra thick,-{Latex[length=10pt, width=10pt]}] (0,0) -- (\ta:1);
  \node[mplorange!60, fill=white, fill opacity=0.88, text opacity=1, inner sep=1.5pt, rounded corners=1.5pt] at (70:0.45) {\LARGE $\hat{y}$};
  \fill[mplgreen!80!black] (5*\s,5*\s) circle (1.4pt);
  \node[mplgreen!80!black, fill=white, fill opacity=0.88, text opacity=1, inner sep=1.2pt, rounded corners=1.5pt, above right=4pt and -1pt, align=center] at (5*\s,5*\s) {\Large solution\\[4pt] \Large $x$};
  \draw[mplpurple, thick, dash pattern=on 3pt off 1.5pt, rotate around={\ta:(\ta:0.95)}] (\ta:0.95) ellipse[x radius=0.08, y radius=0.62];
  \node[mplpurple, fill=white, inner sep=0.8pt] at (0.03,1.19) {\Large $Q$};
  \node[gray!75,fill=white,inner sep=0.6pt] at (-0.935,-0.85) {\Large $\mathbb{Z}^{2r}$};
  \node[fill=white, inner sep=0.6pt] at (-50:1.26) {\large $\norm{x}^2 = 2^k$};
  \node[mplblue!62,fill=white,inner sep=0.5pt, left=4pt] at (65:0.85) {\Large cap};
\end{tikzpicture}
\caption{
The search geometry of Sec.~\ref{sec:problem}, drawn in the 2D plane (the real search runs in $\R^{2r}$, $2r=8$ or $16$).
A feasible candidate is an integer lattice point (green) that lies on the norm shell (Eq.~\eqref{eq:normshell}) \emph{and} inside the cap (Eq.~\eqref{eq:cap}, shaded blue region), the shell region aligned with the target direction $\hat y$.
Black points on the shell satisfy the bilinear conditions (Eq.~\eqref{eq:bilinear-omega}) but lie outside the cap, and gray points on the shell fail them.
The cap is exaggerated for legibility.
\textsc{Form} encloses the cap in the thin ellipsoid $Q$ (dashed), \textsc{LLL} reduces the lattice basis under $Q$, and the \textsc{SE} enumeration walks the lattice points inside it, starting from the rounded cap center $z_c$ (Sec.~\ref{sec:algorithm}).
}
\label{fig:cap-geometry}
\end{figure}
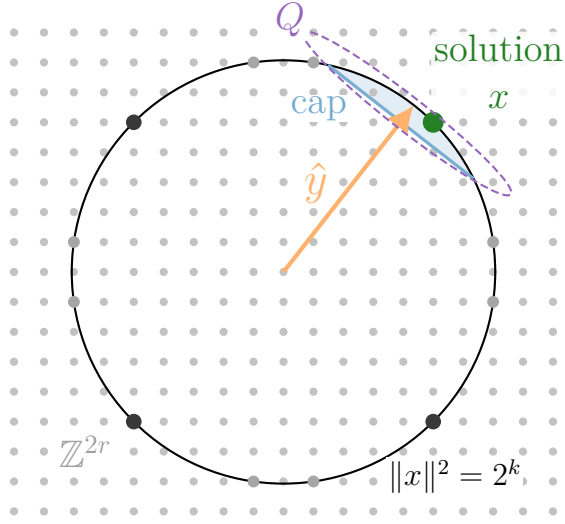

\section{Algorithm}
\label{sec:algorithm}

Both gate sets are handled similarly.
First a depth $k$ is swept and at each depth band $k$, integer vectors satisfying the norm shell, bilinear equations, and cap of Sec.~\ref{sec:problem} are enumerated.
We present the two cases separately to isolate what changes when swapping $T$ with $\sqrt{T}$.
Algorithm~\ref{alg:ct} sketches the Clifford+$T$ procedure of Morisaki~\textit{et al.}~\cite{morisaki2025optimalancillafreeclifford+t}.
Algorithm~\ref{alg:sqrtt} is our contribution.

\subsection{\texorpdfstring{Clifford+$T$}{Clifford+T} synthesis}

The Clifford+$T$ procedure works in $\Z[\omega]$, where a candidate is $8$ integers subject to one bilinear equation.
Algorithm~\ref{alg:ct} keeps only the lattice search that both gate sets share.
The full algorithm of Morisaki~\textit{et al.}, which our implementation follows, sweeps the $\Tcount$ itself rather than the lde, splits each candidate into a short prefix drawn from an enumerated set and a lattice-searched remainder, and reaches odd $\Tcount$s by retargeting $V\gets VT^\dagger$~\cite{morisaki2025optimalancillafreeclifford+t}.
Because the $\Tcount$ is swept in increasing order and each level is enumerated in full, the first feasible circuit has the minimal $\Tcount$.
Table~\ref{tab:doubling} collects the quantities that change when the same template is instantiated for Clifford+$\sqrt{T}$.

\begin{table}[t]
\centering
\caption{
    Moving from Clifford+$T$ to Clifford+$\sqrt{T}$ replaces the ring $\Z[\omega]$ by the larger $\Z[\zeta]$.
    Every relevant size doubles, and one extra consequence is two additional bilinear equations: their number is one fewer than the rank of the conjugation-fixed real subring $\mathcal{R}\cap\R$ (rank $2$ for $\Z[\sqrt2]$, rank $4$ for $\ztr$).
    In either ring the non-Clifford count of an implementable unitary grows linearly with its lde $k$ (for Clifford+$T$, $\Tcount\in\{2k\!-\!2,\,2k\}$), so a deeper $k$ buys a finer mesh at proportionally higher cost.
}
\label{tab:doubling}
\footnotesize\setlength{\tabcolsep}{3pt}
\begin{tabular}{lcc}
\toprule
 & Clifford+$T$ & Clifford+$\sqrt{T}$ \\
\midrule
ring $\mathcal{R}$ & $\Z[\omega]$ & $\Z[\zeta]$ \\
real subring $\mathcal{R}\cap\R$ & $\Z[\sqrt2]$ & $\ztr$ \\
integers per ring element ($r$) & $4$ & $8$ \\
bilinear equations & $1$ & $3$ \\
lattice dimension  & $8$ & $16$ \\
\bottomrule
\end{tabular}
\end{table}

\begin{algfloat}{Clifford+$T$ synthesis (sketch of~\cite{morisaki2025optimalancillafreeclifford+t})}
\label{alg:ct}
\begin{algorithmic}[1]
\Require target $V$, tolerance $\varepsilon$
\Ensure $U_T\in\langle H,S,T\rangle$, $\dd(U_T,V)<\varepsilon$
\State $d \gets \textsc{DetPhase}(V)$ \Comment{even parity ($\Z[\omega]$)}
\For{$k = 0, 1, 2, \dots$} \Comment{shallowest first}
  \State $y \gets \textsc{Align}(V, k, d)$ \Comment{target $y\in\R^{8}$}
  \State $Q \gets \textsc{Form}_8(y, k, \varepsilon)$ \Comment{cap ellipsoid (Fig.~\ref{fig:cap-geometry})}
  \State $B \gets \textsc{LLL}(Q)$ \Comment{reduced basis of $\Z^{8}$}
  \State $z_c \gets \textsc{Round}(B, y, \varepsilon)$ \Comment{cap center}
  \For{$z \in \textsc{SE}(B, z_c)$} \Comment{enumerate near $z_c$}
    \State $x \gets B z$
    \If{$\textsc{Feasible}(x)$} \Comment{Eqs.~\eqref{eq:normshell}--\eqref{eq:cap}}
      \State \Return $U_T \gets \textsc{Decode}_\omega(x, d)$
    \EndIf
  \EndFor
\EndFor
\end{algorithmic}
\end{algfloat}

\subsection{\texorpdfstring{Clifford+$\sqrt{T}$}{Clifford+sqrt(T)} synthesis}

Our procedure follows the same template in the larger ring $\Z[\zeta]$, where a candidate is $16$ integers subject to three bilinear equations (Sec.~\ref{sec:problem}).
The steps marked {\color{blue}$\blacktriangleright$} in Algorithm~\ref{alg:sqrtt} have no counterpart in Algorithm~\ref{alg:ct}; the rest of the body is Algorithm~\ref{alg:ct} with the ring-dependent parameters swapped.

Two of these additions are worth explaining.
First, a pure-$T$ solution sits at a depth $k$ outside the window enumerated by the $\sqrt{T}$ infrastructure.
This means Algorithm~\ref{alg:sqrtt} will not find the optimal pure-$T$ solution directly.
Instead we run Clifford+$T$ synthesis concurrently and store the result so the output never costs more than the Clifford+$T$ optimum.

The second addition is cost minimization.
Whether the first feasible circuit found is also the cheapest depends on the gate set.
For Clifford+$T$ it is, because the $\Tcount$ is swept in increasing order, so the first feasible circuit is already $\Tcount$-optimal.
For Clifford+$\sqrt{T}$ a deeper band can reach the target with a cheaper gate mix, so the first feasible circuit is often not the least costly.
Writing $k_{\mathrm{hit}}$ for that shallowest feasible band, we continue enumerating through band $k_{\mathrm{hit}}+w$ for a small window $w$, keeping the minimum-cost circuit found.
The window is short because the chance of a further saving diminishes as the band grows.

\begin{algfloat}{Clifford+$\sqrt{T}$ synthesis (this work)}
\label{alg:sqrtt}
\begin{algorithmic}[1]
\Require target unitary $V$, tolerance $\varepsilon$, cost-search window $w$ (default $2$), optional deadline $t_{\max}$
\Ensure $U_{\sqrt T}\in\langle H,S,\sqrt{T}\rangle$, $\dd(U_{\sqrt T},V)<\varepsilon$, low cost
\State \dnew $U_{\sqrt T} \gets \textsc{Alg.}\,\ref{alg:ct}(V, \varepsilon)$ \Comment{cost floor}
\ForAll{\dnew $V' \in \{V,\, e^{i\pi/16}V\}$} \Comment{parity branches}
  \State $d \gets \textsc{DetPhase}(V')$ \Comment{App.~\ref{app:recon}}
  \State \dnew $k_{\mathrm{hit}} \gets \textsc{Screen}(V', \varepsilon, d)$ \Comment{first feasible}
  \For{\dnew $k = k_{\mathrm{hit}}, \dots, k_{\mathrm{hit}}+w$} \Comment{cost window}
    \State $y \gets \textsc{Align}(V', k, d)$ \Comment{target $y\in\R^{16}$}
    \State $Q \gets \textsc{Form}_{16}(y, k, \varepsilon)$ \Comment{cap ellipsoid}
    \State $B \gets \textsc{LLL}(Q)$ \Comment{reduced basis of $\Z^{16}$}
    \State $z_c \gets \textsc{Round}(B, y, \varepsilon)$ \Comment{cap center}
    \For{$z \in \textsc{SE}(B, z_c)$} \Comment{enumerate near $z_c$}
      \State \dnew \textbf{if} $t_{\max}$ elapsed, \textbf{break} \Comment{keep best}
      \State $x \gets B z$
      \If{$\textsc{Feasible}(x)$} \Comment{Eqs.~\eqref{eq:normshell}--\eqref{eq:cap}}
        \State $U \gets \textsc{Decode}_\zeta(x, d)$ \Comment{Ref.~\cite{forest2015exactsynthesissinglequbit}}
        \If{$\mathrm{cost}(U) < \mathrm{cost}(U_{\sqrt T})$}
          \State \dnew $U_{\sqrt T} \gets U$
        \EndIf
      \EndIf
    \EndFor
  \EndFor
\EndFor
\State \Return $U_{\sqrt T}$ \Comment{cheapest found, $\le$ Clifford+$T$}
\end{algorithmic}
\end{algfloat}

Both algorithms are built from the same pipeline of subroutines:

\begin{description}[leftmargin=0pt,labelindent=0pt,labelsep=0.5em,itemsep=2pt]
\item[DetPhase] computes the target's determinant (global-phase) class $d$, fixing the reconstruction so the candidate's determinant matches the target's (Appendix~\ref{app:recon}).

\item[Screen] finds $k_{\mathrm{hit}}$, the band where the cost search begins. It reuses the per-band search detailed below, running it from the shallowest band upward and taking $k_{\mathrm{hit}}$ to be the first band whose search returns a feasible point.

\item[Align] embeds the target at depth $k$ as a real vector $y\in\R^{2r}$. This is the direction that a candidate on the norm shell must point toward.

\item[Form] builds the positive-definite matrix $Q$ whose unit ellipsoid encloses the cap (Fig.~\ref{fig:cap-geometry}), so that vectors short under $Q$ are exactly the near-feasible candidates.

\item[LLL] reduces the lattice $\Z^{2r}$ under $Q$~\cite{nguyen2009lllalgorithmquadratic,lenstra1982factoringpolynomialsrational}, producing the basis $B$.
This reduction is what makes subsequent stages output-sensitive.

\item[Round] rounds the cap center into lattice coordinates, returning the integer point $z_c$ from which the enumeration starts.
Good candidates are expected near the center of the cap rather than along its edges, so the enumeration begins there.

\item[SE] the Schnorr--Euchner~\cite{schnorr1994latticebasisreduction} enumeration walks lattice points outward from $z_c$.
Its worst-case cost is exponential in the lattice dimension, but for the dimensions here it is fast in practice, touching far fewer points than exhaustively enumerating the cap.

\item[Feasible] tests each enumerated $x=Bz$ against three conditions: the norm shell $\norm{x}^2=2^k$ (tested first, as it rejects almost all points), the bilinear forms $\Phi_j(x)=0$, and the alignment cap $(\hat y\cdot x)^2\ge\tau$ of Eq.~\eqref{eq:cap}.
It returns true only when all hold.
The first two are exact integer tests, so they do not depend on the conditioning of $Q$.
The cap involves the real target $\hat y$ and is evaluated in arbitrary-precision arithmetic.

\item[Decode] turns the accepted ring element into a gate string by the exact synthesis of Forest~\textit{et al.}~\cite{forest2015exactsynthesissinglequbit}, whose canonical form covers both $\Z[\omega]$ and $\Z[\zeta]$ and uses the fewest $\pi/8$ rotations.
\end{description}

\section{Results}
\label{sec:results}

\subsection{Setup}

For each precision $\varepsilon\in\{10^{-3},\dots,10^{-8}\}$ we draw $500$ Haar-random unitaries.
Each target is synthesized twice to $\dd(U,V)<\varepsilon$ with our open-source implementation \texttt{cyclosynth}: once with the optimal Clifford+$T$ algorithm~\cite{morisaki2025optimalancillafreeclifford+t} and once with our Clifford+$\sqrt{T}$ algorithm.

We report the per-target relative cost 
\begin{equation}
  \rho = \frac{\mathrm{cost}(U_{\sqrt T})}{\mathrm{cost}(U_T)},
  \label{eq:rho}
\end{equation}
where $\rho<1$ means $\sqrt{T}$ is cheaper.
All trials use a parallelized implementation of Algorithms~\ref{alg:ct} and~\ref{alg:sqrtt} run on an Apple M4 Pro with $14$ cores, one target at a time.
Targets are drawn from fixed seeds, and the script \texttt{scripts/reproduce\_paper.sh} in the \texttt{cyclosynth} repository regenerates every figure and table of this section (Data availability).
Every reported circuit is verified after synthesis by reconstructing its unitary and recomputing $\dd(U,V)$.

\subsection{Cost}

The median relative cost $\rho$ is $0.75$--$0.79$. This corresponds to a $21$--$25\%$ median cost reduction against the $\Tcount$-optimal baseline (Table~\ref{tab:cost}).
The comparison is paired, so both gate sets synthesize the same 500 targets.

In absolute terms the mean non-Clifford cost grows linearly in $\log_2(1/\varepsilon)$.
The scaling factor is estimated using a least-squares fit of the mean per-target cost over the six precisions $10^{-3}$ to $10^{-8}$.
The cost of Clifford+$T$ circuits scales roughly as $3.0\log_2(1/\varepsilon)$. 
This matches the $3\log_2(1/\varepsilon)$ lower bound for typical unitaries~\cite{selinger2014efficientclifford+tapproximation}, which the algorithm of Morisaki~\textit{et al.} attains~\cite{morisaki2025optimalancillafreeclifford+t}.
The Clifford+$\sqrt{T}$ cost scales as $2.4\log_2(1/\varepsilon)$, a ${\sim}20\%$ reduction in cost scaling.
The fits are displayed as dashed lines in Fig.~\ref{fig:cost-headline}.

Synthesizing over Clifford+$\sqrt{T}$ is \emph{strictly} cheaper for $99.4$--$100\%$ of targets at every $\varepsilon$, the remainder being ties at $\rho=1$.
No target is ever costlier than Clifford+$T$ by the floor of Algorithm~\ref{alg:sqrtt}.
The $\sqrt{T}$ synthesis algorithm also uses fewer Clifford gates, a median of $25$ versus $45$ at $\varepsilon=10^{-3}$ and $64$ versus $133$ at $10^{-8}$.

\begin{figure*}[t]
\centering
\includegraphics[width=\textwidth]{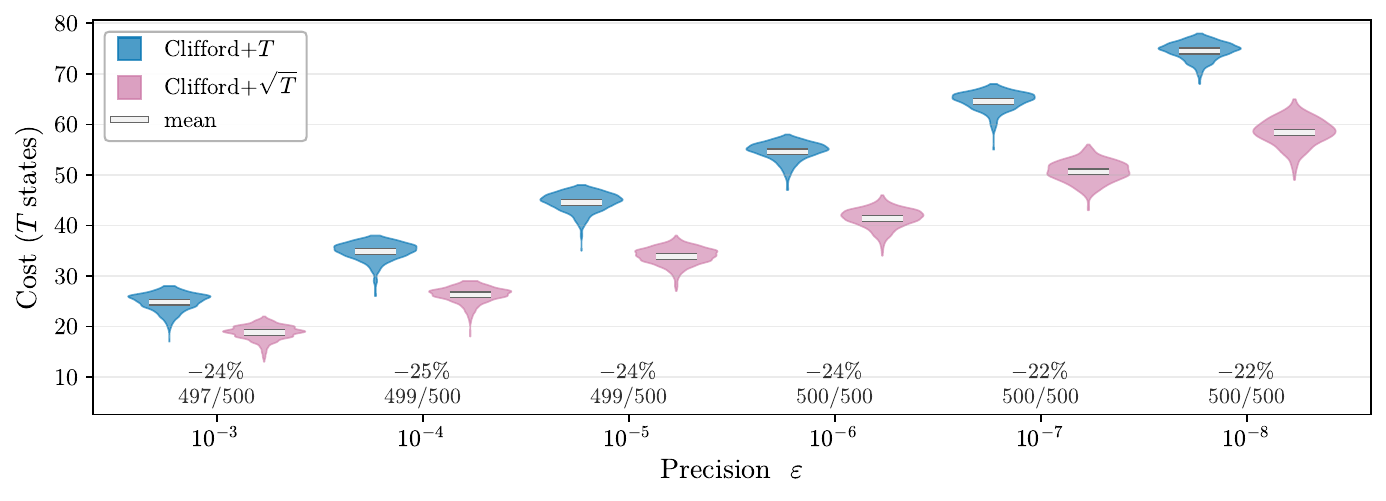}
\caption{Distribution of per-target cost $n_T+3\,n_{\sqrt{T}}$ (Eq.~\eqref{eq:cost}) at each precision: Clifford+$T$ (blue, left) vs.\ Clifford+$\sqrt{T}$ (purple, right), over $500$ Haar-random targets per $\varepsilon$, with the mean marked (light line).
Targets are sampled independently at each $\varepsilon$.
The comparison is paired within each $\varepsilon$ bucket so both gate sets synthesize the same targets.
Annotations give the median cost reduction and the strictly-cheaper count.}
\label{fig:violin}
\end{figure*}

The advantage is largest at loose $\varepsilon$ and narrows slightly as $\varepsilon$ tightens.
We caution that the narrowing is not necessarily intrinsic to the gate sets.
The cost-optimization process explores a bounded band rather than the entire feasible set (Sec.~\ref{sec:algorithm}), so at the finest $\varepsilon$ the returned $\sqrt{T}$ circuits are less likely to be globally cost-optimal.
Figure~\ref{fig:violin} shows the distribution of costs for the $500$ Haar-random general unitaries in $\mathrm{SU}(2)$.

\begin{table}[t]
\centering
\caption{
Cost comparison of Clifford+$\sqrt{T}$ vs.\ optimal Clifford+$T$ over $500$ Haar-random targets per $\varepsilon$.
Cost is the resource cost of Eq.~\eqref{eq:cost} (Sec.~\ref{sec:cost}).
We report the median relative cost $\rho$ (Eq.~\eqref{eq:rho}) per target.
\emph{\% cheaper} is the fraction of targets with $\rho<1$ (the remainder tie at $\rho=1$).
The med cost$_T$ column is the Clifford+$T$ baseline cost (its $T$-count $n_T(U_T)$), the denominator of $\rho$ and the reference for the crossover of Sec.~\ref{sec:discussion}; the $\sqrt{T}$ circuit's own median cost is approximately $\rho$ times it.
The last two columns give the $\sqrt{T}$ circuit's median $T$-type and $\sqrt{T}$-type syllable counts $n_{T}$ and $n_{\sqrt T}$.
}
\label{tab:cost}
\footnotesize\setlength{\tabcolsep}{2.5pt}
\begin{tabular}{lccccc}
\toprule
    $\varepsilon$ & med $\rho$ & \% cheaper & med cost$_T$ & med $n_T$ & med $n_{\sqrt{T}}$ \\
\midrule
$10^{-3}$ & $0.760$ & $99.4\%$ & $25.0$ & $9$ & $3$ \\
$10^{-4}$ & $0.750$ & $99.8\%$ & $35.0$ & $12$ & $5$ \\
$10^{-5}$ & $0.761$ & $99.8\%$ & $45.0$ & $15$ & $6$ \\
$10^{-6}$ & $0.764$ & $100\%$ & $55.0$ & $18$ & $8$ \\
$10^{-7}$ & $0.785$ & $100\%$ & $65.0$ & $18$ & $11$ \\
$10^{-8}$ & $0.784$ & $100\%$ & $75.0$ & $20$ & $13$ \\
\bottomrule
\end{tabular}
\end{table}

\subsection{Comparison with optimal single-axis rotation synthesis}

As an external check we compare against \texttt{gridsynth}~\cite{ross2016optimalancillafreeclifford+t}, the reference Ross--Selinger synthesizer, which is $T$-count-optimal for single-axis rotations.
We match by achieved diamond distance, since \texttt{gridsynth} by default accepts its tolerance as an operator-norm distance.

On $500$ seeded $z$-rotations $R_z(\theta)$ per $\varepsilon$, \texttt{cyclosynth}'s general-purpose Clifford+$T$ search never uses more $T$ gates than \texttt{gridsynth}.
Since the algorithm of Morisaki~\textit{et al.} is $\Tcount$-optimal for any single-qubit unitary, not exceeding the optimal $z$-rotation count is a useful check of our implementation.
On the same $z$-rotations, our Clifford+$\sqrt{T}$ search requires $2.45\log_2(1/\varepsilon)$ $T$ states, compared to $3.09\log_2(1/\varepsilon)$ for Clifford+$T$ (fits as in Sec.~\ref{sec:results}).
We observe that \texttt{cyclosynth}'s Clifford+$T$ $T$-count on general targets is comparable to its $T$-count on single $z$-rotations at the same $\varepsilon$.

On general targets the difference from \texttt{gridsynth} is large (Fig.~\ref{fig:cost-headline}), because \texttt{gridsynth} must use the Euler decomposition $U=R_z(\alpha)R_y(\beta)R_z(\gamma)$.
Each of these single-axis rotations is synthesized independently.
Each rotation is taken to a per-rotation precision ${\sim}\varepsilon/3$, since the diamond distance is subadditive under composition.
We sweep this budget and accept the loosest split whose \emph{composed} circuit is still within $\dd\le\varepsilon$ of $U$.
Across $\varepsilon\in[10^{-3},10^{-8}]$, Clifford+$T$ synthesis~\cite{morisaki2025optimalancillafreeclifford+t} is $3.3$--$3.8\times$ cheaper and Clifford+$\sqrt{T}$ $4.2$--$5.1\times$ cheaper than \texttt{gridsynth} (Fig.~\ref{fig:cost-headline}), quantifying the value of direct general-unitary synthesis over decomposition into rotations.

\subsection{Runtime}

Synthesis is offline, run once per logical operation.
We therefore treat gate cost as the operative metric and synthesis runtime as secondary and tunable.
The cost reduction stage is not free: the $16$-dimensional enumeration, with its arbitrary-precision Gram--Schmidt, is substantially more expensive than the $8$-dimensional Clifford+$T$ search.
The median $\sqrt{T}$ runtime is about $0.6$s at $\varepsilon=10^{-3}$ and $7$s at $\varepsilon=10^{-8}$, a $4$--$13\times$ gap over Clifford+$T$ (Table~\ref{tab:runtime}).

\begin{table}[t]
\centering
\caption{Median per-target synthesis runtime in milliseconds (and $\sqrt{T}$ maximum), $500$ targets per $\varepsilon$, Apple M4 Pro ($14$ cores).}
\label{tab:runtime}
\small\setlength{\tabcolsep}{4pt}
\begin{tabular}{lccc}
\toprule
$\varepsilon$ & Clifford+$T$ & Clifford+$\sqrt{T}$ & $\sqrt{T}$ max \\
\midrule
$10^{-3}$ & $46$ & $560$ & $800$ \\
$10^{-4}$ & $52$ & $551$ & $820$ \\
$10^{-5}$ & $66$ & $826$ & $885$ \\
$10^{-6}$ & $95$ & $1040$ & $1834$ \\
$10^{-7}$ & $187$ & $1981$ & $2225$ \\
$10^{-8}$ & $1831$ & $6583$ & $13867$ \\
\bottomrule
\end{tabular}
\end{table}

\section{Discussion}
\label{sec:discussion}

We have presented the first \emph{direct}, deterministic, ancilla-free approximate-synthesis algorithm for general single-qubit unitaries targeting the Clifford+$\sqrt{T}$ gate set.
Our synthesis algorithm empirically beats the $\Tcount$-optimal Clifford+$T$ construction, with mean cost scaling $2.4\log_2(1/\varepsilon)$ versus $3.0\log_2(1/\varepsilon)$ $T$ states.\\

\begin{figure}[t]
\centering
\includegraphics[width=\columnwidth]{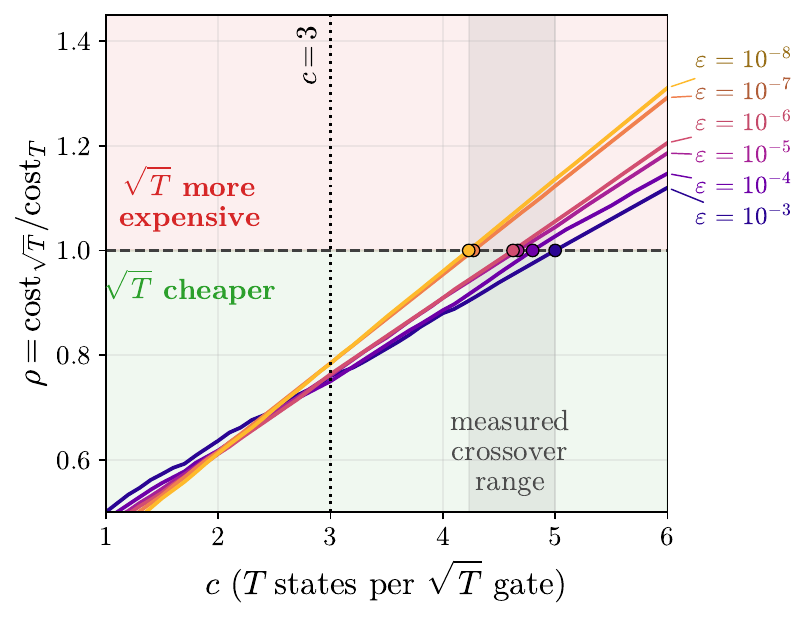}
\caption{
    Sensitivity of the cost comparison to the price of a $\sqrt{T}$ gate.
    The horizontal axis sweeps $c$, the assumed cost of one $\sqrt{T}$ gate in $T$ states; the vertical axis is the relative cost $\rho(c)=\mathrm{cost}(U_{\sqrt T})/\mathrm{cost}(U_T)$, the cost (Eq.~\eqref{eq:cost}) of our Clifford+$\sqrt{T}$ circuit divided by that of the Clifford+$T$ circuit for the same target.
    Each curve traces the median of $\rho(c)$ at the stated precision $\varepsilon$ over the same $500$ Haar-random targets as Fig.~\ref{fig:violin}.
    Wherever a curve lies below $1$ (green region), Clifford+$\sqrt{T}$ is cheaper for the median target.
    A circle marks each curve's crossover $c^\star(\varepsilon)$, the $\sqrt{T}$ price at which the median $\rho$ reaches $1$: below $c^\star$ the $\sqrt{T}$ circuits win, above it the Clifford+$T$ baseline does.
    The gray band spans the crossovers' range across precisions.
    The catalyzed reference price $c=3$ (dotted line) lies below every crossover.
}
\label{fig:csweep}
\end{figure}

\noindent\emph{Robustness to the price of $\sqrt{T}$ gates.}
The $\sqrt{T}$ price $c=3$ (in $T$ states) is an \emph{upper bound} from the catalyzed construction of Sec.~\ref{sec:cost}, and the true price is architecture-dependent.
Figure~\ref{fig:csweep} therefore treats the price as a free parameter and sweeps it.
The $\sqrt{T}$ advantage persists for every price below the crossover $c^\star(\varepsilon)$, where the median relative cost reaches $1$.
The measured crossovers, $c^\star=4.2$ to $5.0$, lie well above the reference at every precision.
Replacing the weight $3$ in Eq.~\eqref{eq:cost} by $c$ and setting $\mathrm{cost}(U_{\sqrt T})=\mathrm{cost}(U_T)$ gives the crossover in terms of the syllable counts,
\begin{equation}
  c^\star=\frac{n_T(U_T)-n_T(U_{\sqrt T})}{n_{\sqrt T}(U_{\sqrt T})},
  \label{eq:cstar}
\end{equation}
evaluated on the two circuits for the same target.
From Table~\ref{tab:cost}, as $\varepsilon\to0$ each count grows at a fixed rate per decade of precision: $n_T(U_T)$ by ${\sim}10$, $n_T(U_{\sqrt T})$ by ${\sim}2.2$, and $n_{\sqrt T}(U_{\sqrt T})$ by ${\sim}2$, so $c^\star\to(10-2.2)/2\approx3.9$.
The reference $c=3$ sits below this limit, well inside the favorable region throughout the measured range.

A cheaper $\sqrt{T}$ preparation protocol only amplifies the relative performance of this synthesis approach.
Because Eq.~\eqref{eq:cost} is linear in syllable counts, $c$ is simply the price of a $\sqrt{T}$ state relative to a $T$ state.
The sweep over $c$ is therefore agnostic to how the states are produced.
Even if no circuit gadget can make $\sqrt{T}$ states more cheaply than the catalyzed construction, cultivation~\cite{gidney2024magicstatecultivation} or distillation~\cite{bravyi2005universalquantumcomputation} primitives that prepare the $\sqrt{T}$ state directly~\cite{chen2026efficientmagicstate} may.

\noindent\emph{Optimality.}
Our results yield a concrete algorithm, but we do not prove that we achieve the gate set's information-theoretic optimum.
The optimality proof of~\cite{morisaki2025optimalancillafreeclifford+t} rests on a relation between $\Tcount$ and lde that is specific to $\Z[\omega]$.
For comparison on $z$-rotations, the Clifford+$\sqrt{T}$ diagonal approximations of Kliuchnikov~\textit{et al.}~\cite{kliuchnikov2023shorterquantumcircuitsa} use about $0.6\log_2(1/\varepsilon)$ $T$ gates and $0.6\log_2(1/\varepsilon)$ $\sqrt{T}$ gates, which we infer from their gate-count and $\Tcount$ fits (their Table~1, where a $\sqrt{T}$ is charged four $T$ states).
Under our cost model that is about $2.4\log_2(1/\varepsilon)$ $T$ states, compared to $2.45\log_2(1/\varepsilon)$ for our general search run on $z$-rotations (Sec.~\ref{sec:results}).

\noindent\emph{Determinism.}
Synthesis need not be deterministic: two standard relaxations trade a deployment resource for lower expected cost.
\emph{Channel mixing} classically randomizes over several deterministic circuits, each individually a coarser approximation, so that only the \emph{average} channel lies within $\dd<\varepsilon$.
It needs no ancilla or measurement, but the accuracy is an ensemble property over repeated runs rather than a guarantee for any single execution.
\emph{Repeat-until-success} (fallback) circuits instead use an ancilla and mid-circuit measurement, repeating a heralded sub-circuit until it succeeds~\cite{paetznick2014repeatuntilsuccessnondeterministicdecomposition,bocharov2015efficientsynthesisuniversal}.
Each accepted output is exactly within $\dd<\varepsilon$, but the depth and gate count are random.

Both relaxations can beat our deterministic circuits in expected cost.
Morisaki \textit{et al.}~\cite{morisaki2025optimalancillafreeclifford+t} pair their deterministic algorithm with a probabilistic synthesis that emits a mixture of Clifford+$T$ circuits, each using at most ${\sim}1.5\log_2(1/\varepsilon)$ $T$ gates for general unitaries, half the deterministic count and below $2.4\log_2(1/\varepsilon)$.
The price is that the accuracy guarantee holds only on average, raising the number of shots needed on the quantum computer.
Kliuchnikov~\textit{et al.}~\cite{kliuchnikov2023shorterquantumcircuitsa} develop mixing and fallback for the $z$-rotation and magnitude sub-problems of their pipeline, reaching ${\sim}0.5\log_2(1/\varepsilon)$ expected $T$ states per $z$-rotation with mixed fallback.
For general Clifford+$T$ unitaries they estimate $7\log_2(1/\varepsilon)$ deterministically and $3.5\log_2(1/\varepsilon)$ with mixing (their Table~2).

Our algorithm emits a single fixed gate sequence meeting $\dd<\varepsilon$ in one execution, with no ancilla, measurement, or classical randomness.
Deterministic, ancilla-free synthesis is thus the least demanding to deploy.
It is also the only option when those resources are unavailable at the point of use or a single static circuit is required.

\noindent\emph{Mixing composes with our circuits.}
These relaxations are complements to the present algorithm.
The mixing construction draws its ensemble from deterministic circuits.
Because mixing quadratically suppresses the error of its members~\cite{campbell2017shortergatesequences,morisaki2025optimalancillafreeclifford+t}, the ensemble need only be synthesized to ${\sim}\sqrt{\varepsilon}$; applied over $\Z[\zeta]$, the mixing construction of Morisaki~\textit{et al.} would therefore be expected to roughly halve the cost of each circuit in the mixture.
We leave the implementation of this to future work.
The same quadratic suppression also extends the precisions that can be reached.
Ensemble members synthesized at $10^{-8}$, the deepest precision our search runs comfortably, mix to an effective channel precision of ${\sim}10^{-16}$.

\noindent\emph{Generalization beyond $\sqrt{T}$.}
The \textsc{Decode} step uses ancilla-free single-qubit exact synthesis, which exists only when the Clifford-cyclotomic group equals the full unitary group over the ring.
This holds for just four roots of unity, $\zeta_n$ with $n\in\{8,12,16,24\}$~\cite{ingalls2019cliffordcyclotomicgroupeulerpoincare}.
In gate terms that is $T$ ($\zeta_8$) and $\sqrt{T}$ ($\zeta_{16}$), plus the non-power-of-two cases $\zeta_{12},\zeta_{24}$.
It \emph{fails} for $T^{1/4}$ ($\zeta_{32}$) and every higher $2^k$-th root.
The lattice can still round a target into the ring, but the rounded element is generically not realizable as an ancilla-free single-qubit circuit.
The lattice enumeration itself is ring-agnostic, but for the higher $2^k$-th roots exact synthesis needs ancillas~\cite{amy2024exactsynthesismultiqubit}.

\noindent\emph{Catalyst dependence.}
The $\sqrt{T}$ cost $c=3$ presumes a prepared $\sqrt{T}$ catalyst state (Sec.~\ref{sec:cost}).
Its ${\sim}6\log_2(1/\varepsilon)$-$T$-state preparation (to precision $\varepsilon^2$) makes the construction worthwhile only when amortized across many rotations.
A catalyst also accumulates error with use and must be discarded once an error is detected~\cite{gidney2019efficientmagicstate}, so the preparation recurs once per catalyst lifetime rather than once per program.
Catalysts for dyadic phase rotations can also be prepared by cultivation~\cite{xu2026cultivatinglogicalcatalysts}.

\noindent\emph{Open directions.}
A $\Z[\zeta]$ counterpart of the $\Tcount$--lde relation behind the optimality proof of~\cite{morisaki2025optimalancillafreeclifford+t} would settle whether our circuits are optimal rather than merely cheaper.
The other exactly-synthesizable Clifford-cyclotomic levels, the $\zeta_{12}$ and $\zeta_{24}$ gate sets~\cite{ingalls2019cliffordcyclotomicgroupeulerpoincare}, invite the same lattice treatment.
Cheaper $\sqrt{T}$ state preparation, whether by circuit gadget or by direct cultivation or distillation, would widen the advantage.
Finally, ancillas and measurement (e.g.\ repeat-until-success circuits) outperform ancilla-free $z$-rotation synthesis~\cite{paetznick2014repeatuntilsuccessnondeterministicdecomposition,bocharov2015efficientsynthesisuniversal}.
Adapting the present lattice search to emit such circuits is a natural route to lower cost.

\section*{Acknowledgments}
This work was supported by the DOE under contract DE-AC02-05CH11231 through the Office of Advanced Scientific Computing Research (ASCR) Quantum Algorithms Team and Accelerated Research in Quantum Computing programs.
This work was also supported by the NSF Quantum Leap Challenge Institute for Quantum Computation (CIQC) under Award No. 2016245.

\section*{Author contributions}
M.W. and J.W.K. developed the algorithm.
M.W. implemented the algorithm and performed the numerical experiments.
J.Ka. contributed to the theoretical development.
J.Ku. and C.I. supervised the project and acquired funding.
Large language models were used to assist in preparing the text of the manuscript and to profile and modify the cyclosynth code.
All AI-assisted output was reviewed, tested, and verified by the authors, who take full responsibility for the work.

\section*{Data availability}
The synthesizer and the scripts that reproduce the reported numbers are available at \url{https://github.com/mtweiden/cyclosynth}.

\appendix

\section{Embedding lattice coordinates}
\label{app:embedding}

We give the construction for the Clifford+$\sqrt{T}$ ring $\Z[\zeta]$; the Clifford+$T$ ring $\Z[\omega]$ is the smaller, analogous case of Morisaki~\textit{et al.}~\cite{morisaki2025optimalancillafreeclifford+t}.

\emph{Integer coordinates.}
Each entry of a column is written in the integral power basis $\{1,\zeta,\dots,\zeta^{7}\}$ with integer coefficients,
\begin{equation}
  u_1=\sum_{j=0}^{7} a_j\,\zeta^{j},\qquad
  u_2=\sum_{j=0}^{7} b_j\,\zeta^{j},\qquad a_j,b_j\in\Z,
\end{equation}
so the pair $(u_1,u_2)$ is recorded as the lattice point
\begin{equation}
  x=(a_0,\dots,a_7,\,b_0,\dots,b_7)\in\Z^{16}.
  \label{eq:coords}
\end{equation}
This is the lattice $\Z^{16}$ enumerated in Appendices~\ref{app:diamond}--\ref{app:bilinear}.

\emph{Automorphisms.}
The geometry the search uses lives not on the integer coordinates $x$ directly but on their image under the Galois automorphisms $\sigma_\ell:\zeta\mapsto\zeta^{\ell}$, $\ell\in\Z_{16}^\times$ (the units of $\Z_{16}$, i.e.\ the odd residues).
Complex conjugation is $\sigma_{15}$, and it leaves every modulus $|\sigma_\ell(u)|$ fixed, so the eight automorphisms collapse to \emph{four} inequivalent ones: the identity $\sigma_1$ (the literal complex value at $\zeta=e^{2\pi i/16}$) together with $\sigma_3,\sigma_5,\sigma_7$, which generalize the single ``bullet'' automorphism of Morisaki~\textit{et al.}; for $\Z[\omega]$ there are only \emph{two} inequivalent automorphisms, the identity and the bullet.
The embedding $\Sigma$ stacks the four images of the column, one row per automorphism,
\begin{equation}
  \Sigma x=
  \begin{pmatrix}
    \sigma_1(u_1) & \sigma_1(u_2)\\
    \sigma_3(u_1) & \sigma_3(u_2)\\
    \sigma_5(u_1) & \sigma_5(u_2)\\
    \sigma_7(u_1) & \sigma_7(u_2)
  \end{pmatrix},
  \label{eq:sigma}
\end{equation}
where $\sigma_\ell(u_i)=\sum_{j=0}^{7} (u_i)_j\,e^{\,i\ell j\pi/8}$, with $(u_1)_j=a_j$ and $(u_2)_j=b_j$.
Each entry $\sigma_\ell(u_i)\in\C$ is read as a real pair $(\Re,\Im)$, so the four rows (one per automorphism, contributing the four coordinates $\Re\sigma_\ell(u_1),\Im\sigma_\ell(u_1),\Re\sigma_\ell(u_2),\Im\sigma_\ell(u_2)$) form the embedded vector $\Sigma x\in\R^{16}$.
The $16\times16$ matrix $\Sigma$ is invertible, so the integer lattice point $x$ and its embedding $\Sigma x$ carry the same data; the search enumerates the integer $x$.
The explicit $\Z[\omega]$ matrix in this form (identity stacked with bullet) is given by Morisaki~\textit{et al.}~\cite{morisaki2025optimalancillafreeclifford+t}.

Two facts about $\Sigma$ carry into the integer coordinates.
It is orthogonal up to scale, $\Sigma^\top\Sigma=4\,I_{16}$, so $x$ inherits the ordinary Euclidean inner product; and the target enters only through the identity row $\sigma_1$.
The closeness condition is therefore a spherical cap around the $\sigma_1$-image of the target (Appendix~\ref{app:diamond}), and column unitarity becomes a Euclidean norm shell together with three bilinear constraints (Appendix~\ref{app:bilinear}).

\section{Diamond distance}
\label{app:diamond}

For two single-qubit unitaries $U,V$, the diamond distance between their channels has the closed form~\cite{morisaki2025optimalancillafreeclifford+t,kliuchnikov2023shorterquantumcircuitsa}
\begin{equation}
  \dd(U,V)^2 = 1-\tfrac14\,\bigl|\mathrm{tr}(U^\dagger V)\bigr|^2 .
  \label{eq:diamond}
\end{equation}

To reach the cap condition we reduce Eq.~\eqref{eq:diamond} to a condition on the first columns.
If both unitaries are written as in Eq.~\eqref{eq:param} with first column of $U$ as $(u_1,u_2)$ and $V$ with first column $(v_1,v_2)$, then the trace is real:
\begin{equation}
  \mathrm{tr}(U^\dagger V) = 2\,\Re\!\bigl(u_1\bar v_1 + u_2\bar v_2\bigr),
  \label{eq:trace}
\end{equation}
so Eq.~\eqref{eq:diamond} becomes $\dd(U,V)^2 = 1-\Re\!\bigl(u_1\bar v_1 + u_2\bar v_2\bigr)^2$.

Passing to lattice coordinates (Appendix~\ref{app:embedding}), write the candidate column as $x/\sqrt2^{\,k}$, where $x\in\Z^{16}$ holds the coordinates of $(u_1,u_2)$ and $\norm{x}^2=2^k$ (Eq.~\eqref{eq:normshell}), and let $\hat y$ be the target column carried into the lattice coordinates through the $\sigma_1$ rows of the embedding in Eq.~\eqref{eq:sigma}; those rows scale lengths by $2$, so $\norm{\hat y}=2$.
Then Eq.~\eqref{eq:trace} is twice the Euclidean inner product $(\hat y\cdot x)/\sqrt2^{\,k}$, so $\dd(U,V)\le\varepsilon$ becomes exactly
\begin{equation*}
  (\hat y\cdot x)^2 \;\ge\; 2^k(1-\varepsilon^2) \;=\; \tau(k,\varepsilon),
\end{equation*}
the cap condition of Eq.~\eqref{eq:cap}.
The derivation assumes both unitaries are of the determinant-one form of Eq.~\eqref{eq:param}. The $d=1$ branch of Algorithm~\ref{alg:sqrtt} reduces to the same computation. Its reconstruction (Eq.~\eqref{eq:recon-app}) factors as $U=e^{i\pi/16}U'$ with $U'$ of the form of Eq.~\eqref{eq:param}, and its target is $e^{i\pi/16}V$, so the common phase cancels in $U^\dagger V$.

\section{Deriving the three bilinear forms}
\label{app:bilinear}

A candidate column is $(u_1,u_2)/\sqrt2^{\,k}$ with $u_1,u_2\in\Z[\zeta]$, collected into the lattice point $x\in\Z^{16}$ by the coordinate map in Eq.~\eqref{eq:coords}.
Column unitarity is
\begin{equation}
  u_1\bar{u}_1 + u_2\bar{u}_2 = 2^k,
  \label{eq:unitarity}
\end{equation}
where $\bar{u}$ is complex conjugation $\sigma_{15}$ (Appendix~\ref{app:embedding}).

The derivation is just bookkeeping on Eq.~\eqref{eq:unitarity}.
Conjugation fixes exactly the totally-real subring $\Z[\zeta+\zeta^{-1}]=\Z[\sqrt{2+\sqrt2}]$, so each product $u\bar{u}$, and hence the whole left-hand side, lives in that subring.
The subring has rank $4$ over $\Z$ (Table~\ref{tab:doubling}), so expanding the left-hand side in a $\Z$-basis of it produces \emph{four} integer components.
The right-hand side $2^k$ is an ordinary integer, contributing only to the rational component; equating components therefore gives one equation per basis element:
the rational component is the norm shell $\norm{x}^2=2^k$, and the other three must vanish,
\begin{equation}
  \Phi_j(x) = \beta_j(u_1) + \beta_j(u_2) = 0, \qquad j=1,2,3.
  \label{eq:phi}
\end{equation}
Carrying out the multiplication $u\bar{u}$ and reading off the coefficient of each non-rational basis element gives the three forms, in the argument's coordinates $(c_0,\dots,c_7)$ (so $\Phi_j(x)$ in Eq.~\eqref{eq:phi} takes $c=a$ for $u_1$ and $c=b$ for $u_2$, per Eq.~\eqref{eq:coords}):
\begin{align}
  \beta_1(u) &= c_0c_1+c_1c_2+\dots+c_6c_7 - c_0c_7, \nonumber\\
  \beta_2(u) &= c_0c_2+c_1c_3+\dots+c_5c_7 - c_0c_6 - c_1c_7, \nonumber\\
  \beta_3(u) &= c_0c_3+c_1c_4+\dots+c_4c_7 \nonumber\\
             &\qquad - c_0c_5 - c_1c_6 - c_2c_7, \nonumber
\end{align}
the minus signs coming from $\zeta^{8}=-1$.
As a check on the search pipeline, brute-force enumeration of the shells $\norm{x}^2=2^k$ for $k\le4$ confirms that the points with $\Phi_1=\Phi_2=\Phi_3=0$ are exactly the unitary columns realizable in $\Z[\zeta]$.

\section{Reconstruction from the determinant phase}
\label{app:recon}

Synthesis is phase-agnostic, so the determinant matters only up to global phase; what survives is its parity, the $\sqrt{T}$-count mod 2.
For Clifford+$\sqrt{T}$ the determinant is a power $\zeta^{d}$ of $\zeta=e^{2\pi i/16}$ ($d\in\Z_{16}$), and only its parity $d\bmod 2$ matters.
An even determinant comes from an even $\sqrt{T}$-count and an odd one from an odd $\sqrt{T}$-count.
The target fixes neither parity, so Algorithm~\ref{alg:sqrtt} runs one branch per parity, on $V$ and on $e^{i\pi/16}V$, and keeps the cheaper result. Each branch reconstructs the implementable unitary from its first column $(u_1,u_2)$, with $d\in\{0,1\}$ its parity, as
\begin{equation}
  U = \frac{1}{\sqrt2^{\,k}}
  \begin{pmatrix} u_1 & \zeta^{\,d}(-\bar{u}_2)\\[2pt] u_2 & \zeta^{\,d}\,\bar{u}_1\end{pmatrix}.
  \label{eq:recon-app}
\end{equation}
The factor $\zeta^{d}$ in the second column carries the determinant phase.
Without the odd branch the reconstruction realizes only even parity, leaving odd-$\sqrt{T}$-count targets such as $H\sqrt{T}H$ unrealizable by this reconstruction.

\bibliographystyle{quantum}
\bibliography{references}

\end{document}